\documentclass{aa}

\usepackage{epsfig}     
\usepackage{graphicx,color}     
\usepackage{amssymb}            
\usepackage{url}                
\usepackage{amsmath}            
\usepackage{rotating}                   
\usepackage{float}                      
\usepackage{textcomp}
\usepackage{epstopdf}
\usepackage{dcolumn}
\usepackage{times}
\usepackage{tabularx}
\usepackage{hyperref}
\usepackage[normalem]{ulem}
\hypersetup{
    colorlinks,
    citecolor=blue,
    filecolor=blue,
    linkcolor=blue,
    urlcolor=blue,
    menucolor=black}
\usepackage{soul} 
\usepackage[english]{babel}
\usepackage{booktabs}
\usepackage{xspace}
\usepackage{gensymb}
\usepackage{arevmath}
\usepackage{orcidlink}

\newcommand{\lastpaper}{Paper I}
\newcommand{\hrieuv}{HRI\textsubscript{EUV}\xspace}
\newcommand{\aia}[1]{#1~{\AA}\xspace}
\newcommand{\slithri}[1]{~S$_{\tiny \mathrm{HRI}}^{#1}$\xspace}
\newcommand{\slitaia}[2][2]{~S$_{\tiny \mathrm{#2}}^{#1}$\xspace}

\begin{document}

\title{ Anomalous cross-field motions of solar coronal loops}

\author{
Sudip~Mandal\inst{1}\orcidlink{0000-0002-7762-5629},
Hardi~Peter\inst{1,2}\orcidlink{0000-0001-9921-0937},
James~A.~Klimchuk\inst{3}\orcidlink{0000-0003-2255-0305},
\and
Lakshmi~Pradeep~Chitta\inst{1}\orcidlink{0000-0002-9270-6785}}

\institute{
Max Planck Institute for Solar System Research, Justus-von-Liebig-Weg 3, 37077, G{\"o}ttingen, Germany \\
\email{smandal.solar@gmail.com}
\and
Institut f{\"u}r Sonnenphysik (KIS), Georges-Köhler-Allee 401a
79110 Freiburg, Germany
\and
NASA Goddard Space Flight Center, USA
}

\abstract{
 Here, we present several examples of unusual evolutionary patterns in solar coronal loops that resemble cross-field drift motions. These loops were simultaneously observed from two vantage points by two different spacecraft: the High-Resolution Imager (\hrieuv) of the Extreme Ultraviolet Imager aboard the Solar Orbiter and the Atmospheric Imaging Assembly (AIA) aboard the Solar Dynamics Observatory. Across all these events, a recurring pattern is observed: Initially, a thin, strand-like structure detaches and shifts several megameters (Mm) away from a main or parent loop. During this period, the parent loop remains intact in its original position. After a few minutes, the shifted strand reverses its direction and returns to the location of the parent loop. Key features of this `split-drift' type evolution are: (i) the presence of kink oscillations in the loops before and after the split events, (ii) a sudden split motion at about 30\,km\,s$^{-1}$, with additional slow drifts, either away from or back to the parent loops, at around 5\,km\,s$^{-1}$. Co-temporal photospheric magnetic field data obtained from the Helioseismic and Magnetic Imager (HMI) reveal that during such split-drift evolution, one of the loop points in the photosphere moves back and forth between nearby magnetic polarities. While the exact cause of this `split-drift' phenomenon is still unclear, the consistent patterns observed in its characteristics indicate that there may be a broader physical mechanism at play. This underscores the need for further investigation through both observational studies and numerical simulations.
}

   \titlerunning{Anomalous cross-field motions of solar coronal loops}
   \authorrunning{Sudip Mandal et al.}
   \maketitle
 
\section{Introduction} \label{sec:intro}
The solar corona, the Sun’s outermost layer, is an extremely hot (exceeding a million Kelvin), magnetically dominated low-dense, and highly dynamic region. This low plasma-$\beta$ environment supports a variety of solar features, with coronal loops being especially prominent due to their bright, arch-like shapes in the extreme ultraviolet (EUV) and X-ray wavelengths. These loops trace magnetic field lines, making them essential for studying magnetic field interactions and evolution in the corona. For a thorough review of coronal loop observations and models, see \citet{2014LRSP...11....4R}.

Despite several decades of research on coronal loops, many aspects of these features remain poorly understood. For example: (i) the substructure of loops and its contribution to heating processes \citep{2013A&A...556A.104P}; (ii) the observed constancy in loop width with height, which conflicts with predictions from field extrapolation models \citep{1992PASJ...44L.181K,2000SoPh..193...53K,2012A&A...548A...1P}; (iii) the reason for loops’ apparent circular cross-sectional shape \citep{2020ApJ...900..167K}; and (iv) identifying the heating mechanisms that sustain loop temperatures at million-Kelvin levels \citep{2006SoPh..234...41K}. Because coronal emission is optically thin, observing loops from multiple perspectives can be instrumental in improving estimates of their geometry and dynamics \citep{2007ApJ...671L.205F,2008ApJ...680.1477A,2008ApJ...679..827A,2021ApJ...913...56M}. For instance, \citet{2021ApJ...913...56M} analyzed data from the Solar Dynamics Observatory \cite[SDO;][]{2012SoPh..275....3P} and Solar TErrestrial RElations Observatory \cite[STEREO;][]{2008SSRv..136....5K}, tracking several loops from two viewpoints. They observed that loop cross-sections either showed no correlation or, in some cases, anticorrelations between the instruments. One of the ways to enhance these analyses is to use coronal images with improved spatiotemporal resolution.

This was attempted in a recent study by \citet{2024A&A...682L...9M} (hereafter \lastpaper) who examined a particular loop observed simultaneously with the SDO and the Solar Orbiter spacecraft \citep{2020A&A...642A...1M}, which were separated by an angle of 43\textdegree. The EUV images from the High Resolution Imager (\hrieuv) on the Extreme Ultraviolet Imager \cite[EUI;][]{2020A&A...642A...8R} provided four times the spatial resolution of SDO's Atmospheric Imaging Assembly~\cite[AIA;][]{2012SoPh..275...17L}. Two key findings were presented: (i) \hrieuv’s high-resolution images revealed finer details compared to AIA, but the overall loop morphology, such as cross-sectional shape, remained largely unchanged, supporting the hypothesis of a circular cross-section; and (ii) the loop exhibited complex dynamics, including a split in the main loop into two segments that subsequently moved apart from the original loop (see Fig. 2 of \lastpaper).
This latter result is particularly exciting because, in a low plasma-$\beta$ environment, the plasma is line-tied to the magnetic field, which restricts motions across the field. However, \lastpaper\ only presents a single event of these unusual loop dynamics and provides only a basic overview of it.

\begin{figure*}[!ht]
\centering
\includegraphics[width=0.98\textwidth,clip,trim=0cm 0.5cm 0cm 0.2cm]{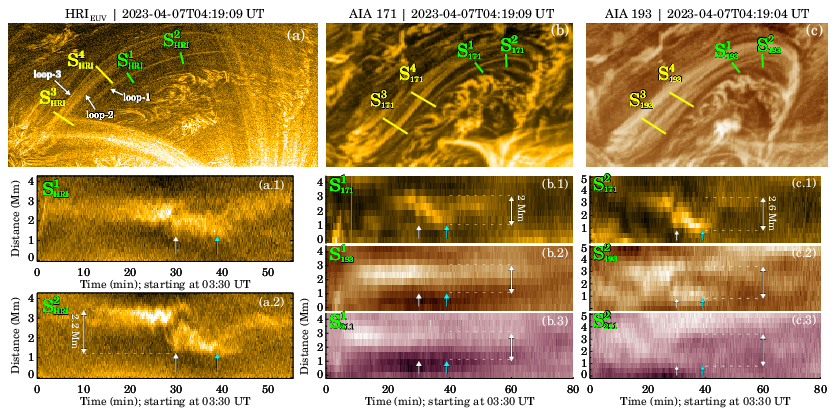}
\includegraphics[width=0.98\textwidth,clip,trim=0cm 0.5cm 0cm 3.1cm]{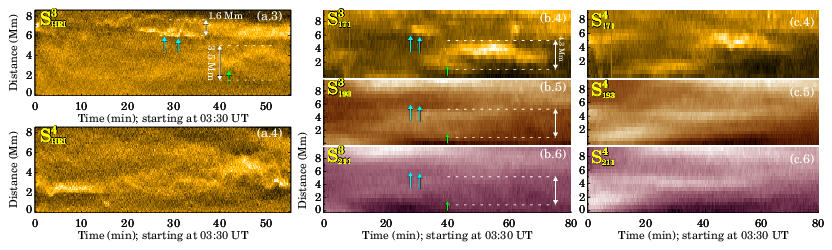}
 
 \caption{Examples of atypical loop evolution events. Panel a presents the \hrieuv image with two sets of yellow and green lines outlining the locations of the artificial slits that were used to derive the space-time (x-t) maps shown in panels a.1 to a.4. The white, blue, and green arrows in these x-t maps mark the times when split-drift type events are observed. The three loop structures where these events occur are highlighted by arrows in panel-a. Panels b and c present the co-temporal images recorded in AIA \aia{171} and \aia{193} passbands, while the x-t maps derived using these data are shown in panels b.1 to c.6. The white, blue, and green arrows on all of these x-t maps mark the identical timestamps as of their \hrieuv counterparts. In b.4 to b.6, the green arrow indicates the only event captured by \slitaia[3]{171}. The other two events on this map, marked by the blue arrows, are only recorded in the \slithri{3} map in a.3. The double-sided arrows in the x-t maps outline the drift separation distance as judged visually. To highlight the loop structures, the context images in panels a, b, and c were created by applying an one-minute running average to the respective datasets.
 Events movies are available via \href{https://drive.google.com/file/d/1ojpZ8ul32N7cqaU8s9_dCdNBvbYpMTou/view?usp=sharing}{EUI\_Slit2\_movie}; \href{https://drive.google.com/file/d/1kxY4fau6NuV6AHpUTn5NuujHXM72rLY5/view?usp=sharing}{AIA\_Slit3\_movie}; \href{https://drive.google.com/file/d/1c22D-61tGqsB9XePXv9C38SiKh8HUIK_/view?usp=sharing}{EUI\_Slit3\_movie}.
 }
\label{fig:context_loop1}
\end{figure*}

In this paper, we revisit the same dataset as \lastpaper, presenting additional events in nearby loops showing similar atypical cross-field motions. We also investigate the relationship between photospheric magnetic field evolution and the unusual dynamics of the loop to gain insight into the drivers behind such loop evolution.


\section{Data} \label{sec:data}

We revisit the same dataset used in Paper I. The observations were taken on April 7, 2023. The \hrieuv dataset\footnote{Publicly available via the SolO/EUI Data Release 6.0 \citep{euidatarelease6}.} was captured using the 174~{\AA} bandpass, with a cadence of 10 seconds and lasted for one hour. It has a plate scale of 108 km per pixel. In addition, we utilized multi-wavelength AIA/SDO data that covers the \hrieuv sequence and extends beyond it. Specifically, we analyzed AIA data from the 171~{\AA}, 193~{\AA}, and 211~{\AA} passbands, each with a cadence of 12 seconds and a plate scale of 435~km per pixel. During this observation period, the angle between SDO and Solar Orbiter was 43\textdegree. Moreover, this study also utilizes line-of-sight (LOS) magnetograms from the Helioseismic and Magnetic Imager (HMI) \citep{2012SoPh..275..207S}, which is also onboard SDO. Additionally, when comparing the SDO and Solar Orbiter data, we corrected for the time difference between the two spacecraft, noting that Solar Orbiter was 0.3 AU away from the Sun while SDO was 1 AU away. All time stamps mentioned are recorded as measured at Earth.

\begin{table*}[!ht]
\caption{Details of `split-drift' events as measured using the \hrieuv data.}  
\label{table}
\begin{center}
\centering
   \begin{tabular} {@{}ccccccc@{}}
         \hline

   Event count & Slit name & Start time & Split speed & Drift speed & Drift distance & Occurrence of oscillations  \\
               &           & (UT)       & (km.s$^{-1}$)& (km.s$^{-1}$)& (Mm) & (before or after)  \\

     \hline
       \rule{0pt}{2ex} 1 & \slithri{2}       & 04:00   & 38 & 4 & 2.2 & both \\
       \\
                       2 & \slithri{2}       & 04:09   & -  & 3 & 2.0 & after (inconclusive)  \\
       \\
                       3 & \slithri{3}       & 03:58   & - &  6  & 0.8 & Inconclusive  \\
       \\
                       4 & \slithri{3}       & 04:01   & 32 &  -  & 1.6 & after  \\
       \\
                       5 & \slithri{3}       & 04:12   & -  &  6  & 3.5 & after (through AIA) \\

  \hline
\end{tabular}
\end{center}
\end{table*}

\section{Results}

\subsection{The common `split-drift' pattern}

We have observed a total of five unusual loop-evolution events (see Fig.~\ref{fig:context_loop1}). From all these events, a common pattern arises: First, a thin strand-like structure separates from the main or parent loop and moves a few Mm to the sideways of the parent loop. The parent loop continues to exist at its original location. After a few minutes, the shifted strand then shows movement in the opposite direction compared to its original movement direction and returns to the location of the parent loop. We refer to this whole evolution sequence as a `split-drift' event. In almost all cases, we find signatures of transverse oscillation, either before or after the thread separation.

While the loops appear visually similar in both \hrieuv and AIA \aia{171} images, subtle differences emerge due to the line-of-sight integration effects, owing to the 43$\degree$ separation angle between the two spacecraft during this observation. This discrepancy is especially noticeable near the footpoints, where loop curvature enhances the effect. Despite this, superior resolution of \hrieuv allows a clearer view of the loops and their evolution compared to AIA. However, the \hrieuv data are available in only one passband (174~{\AA}, sampling plasma at 1~MK), whereas AIA spans a broader temperature range (0.6~MK to 10~MK) with six EUV passbands. Most loops of our interest are not visible in the majority of AIA passbands, so we focus our analysis on the 171{\AA}, 193~{\AA}, and 211~{\AA} channels, where the loops can still be reliably identified.

\subsection{The multi-wavelength aspect}
We first focus on examining the multi-wavelength signatures of these events as well as signatures of such evolution at different positions along the length of these loops. 

Slits \slithri{1} and \slithri{2} cover the lowermost loop (loop-1 in Fig.~\ref{fig:context_loop1}a) and, the two \hrieuv x-t maps derived from these two different slits are shown in panels~\ref{fig:context_loop1}a.1 and a.2. The reason for selecting these two specific slit locations is to investigate the extent of the effect of a split-drift event along the length of a loop, while ensuring that the signal remains adequate at both slit locations. These x-t maps show similar trends of evolution, although the motions are less prominent in the \slithri{1} slit. This is expected as this particular slit is placed slightly further away from the loop top (adjudged visually). The same loop system is also visible in the AIA images and the evolution captured in the two \aia{171} x-t maps\footnote{Note that the position of the AIA slits appears different from those of the \hrieuv due to the distinctly different vantage points of the two spacecraft.} (panels~\ref{fig:context_loop1}b.1,~\ref{fig:context_loop1}c.1), appears similar to the \hrieuv maps. However, they appear completely different when the x-t maps of the \aia{193} (panels~\ref{fig:context_loop1}b.2,~\ref{fig:context_loop1}c.2) and \aia{211} (panels~\ref{fig:context_loop1}b.3,~\ref{fig:context_loop1}c.3) were examined. For example, the \slitaia[1]{193}
shows no noticeable shift of the loop but rather a thick, bright structure staying at the same location throughout the time. The \slitaia[1]{211} map also captures a similar pattern. On the other hand, clear signatures of movement are seen in the \slitaia[2]{193} and \slitaia[2]{211}maps that cover the same loop at a different location, making the multiwavelength evolution difficult to explain. Additionally, the \aia{171} maps reveal that around $\rm{t}$=40~min, the separated loop (or the thread) retracts back to its original position from where the initial movement started. However, no such trend exists in the \aia{193} and \aia{211} maps. Lastly, the initial split in \slitaia[2]{171} map (highlighted by the white arrow), appears approximately a minute later than that of the \slitaia[2]{193} map.

The loops next to the loop-1 also show similar `split-drift' evolution. In fact, multiple of such instances are captured in the \slithri{3} map (panel~\ref{fig:context_loop1}a.3). The overall evolution appear similar to that of the previous case. However, as we move towards the loop top the splitting signatures become difficult to isolate from background structures as revealed in the \slithri{4} map (panel~\ref{fig:context_loop1}a.4). It also noteworthy that the two instances, highlighted by the blue arrows in panel~\ref{fig:context_loop1}a.3, happens simultaneously with the split event seen in loop-1 (highlighted by the white arrow in panel~\ref{fig:context_loop1}a.1). 
The AIA data, however, could resolve only one (highlighted with the green arrow) out of the three cases. The \slitaia[3]{171} map captured clear `split-drift' of the loop starting from t=39 min. Unlike loop-1, the evolution captured in \aia{193} and \aia{211} passbands\footnote{The signal in \aia{211} is significantly weak and the loop can only be identified in hindsight i.e., only after looking at the \aia{171} and \aia{193} x-t maps.} is similar to that of the \aia{171} map. On the other hand, the S$^4$ x-t maps show similar signatures to that of the S$^1$ maps from loop-1 that is the \slitaia[4]{171} map shows the loop moving while the co-temporal \slitaia[4]{193} and \slitaia[4]{211} maps rather show no movements.

\begin{figure*}
\centering
\includegraphics[width=0.98\textwidth,clip,trim=0cm 0cm 0cm 0cm]{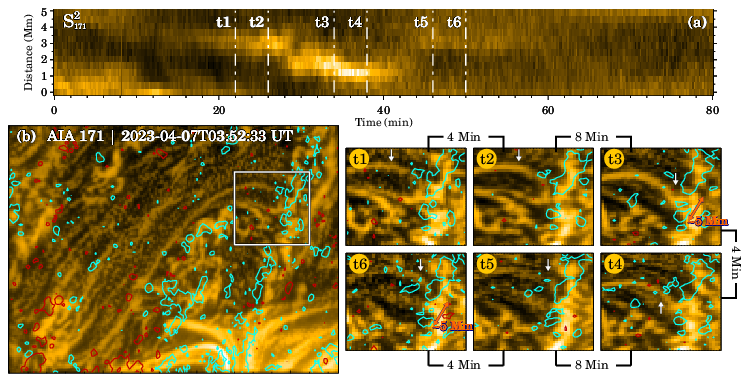}
 \caption{Associations with photospheric magnetic field. Panel a presents the x-t map from \slitaia[2]{171}, while panel b shows a sub-section of the AIA \aia{171} image that encompasses the loop system studied in this paper.  The cyan and red contours overlaid on panel b represent boundaries of $\pm$20~G, derived from the co-temporal HMI LOS magnetogram. The white rectangle in panel b outlines a sub-field of view (FOV) that contains the west-side footpoint region of the loop system. The six side panels next to panel b display snapshots of the sub-FOV at six different times, labeled from t1 to t6. These six timestamps are highlighted with dash-dot vertical lines on the x-t map in panel a. The white arrows in panels t1 to t6 assist in identifying the loop of interest. The back-and-forth movement of the loop footpoint is better visualized in the animation available via \href{https://drive.google.com/file/d/1gUldR6l1GMw5oj0OFeI-3_MibGKGvYzh/view?usp=sharing}{this link.} }
\label{fig:hmi_171}
\end{figure*}
\subsection{The magnetic field underneath}\label{sec:magnetic_field}
Often the dynamics at the coronal heights are a response to the changes in the magnetic field at the photospheric height. In this section, we will investigate this relationship. Polarimetric and Helioseismic Imager  \citep[PHI,][]{2020A&A...642A..11S}, the magnetometer onboard the Solar Orbiter was capturing only a portion of the EUI field of view, and unfortunately, the loops we are interested in lie outside of it. Therefore, we utilized the line-of-sight (LOS) magnetogram from the HMI for our analysis. The loop system is located on a relatively weak magnetic patch, away from the active region (see Appendix; Fig.~\ref{fig:appendix_context}). The footpoints on the east side are anchored in a positive polarity patch (indicated by the red contours), while the west side footpoints are situated in a negative polarity patch (shown by the cyan contours).

The animation associated with Fig.~\ref{fig:hmi_171} reveals that in the coronal images, the west-side footpoint of the loop exhibits significant movement during the split-drift motion of the loop. To investigate this in detail, we selected six time instances (t1 to t6, as marked in Fig.~\ref{fig:hmi_171}) that cover the entire sequence of the event, including the periods before, during, and after the loop split. Close-ups of this footpoint region, with magnetic field contours overlaid, are shown in the panels beside Fig.~\ref{fig:hmi_171}. Through these images, we observe that this loop footpoint seems to jump back and forth between two negative polarity patches during the split event. For instance, at time t1 (before the split), the footpoint is rooted in the larger negative polarity patch near the top of the inset image. By time t3 (after the split), it has moved to the lower patch. Subsequently, at time t5 (and at t6), when the loop has retracted to its initial location, the footpoint also returns to the upper patch. These two polarity patches are separated by a distance of 5~Mm. In comparison, the loop top, according to Fig~\ref{fig:context_loop1}c.1 x-t map, only moved about 3~Mm. At the same time, we observed that the magnetic patches themselves do not move around appreciably over the course of the loop evolution.

Linking loop footpoints identified in EUV images to the photospheric magnetic patches recorded in magnetograms is not always straightforward \citep{2017ApJS..229....4C,2024ApJ...970..147J}. Factors such as low-lying foreground or background structures in the EUV images, loop geometry and magnetic field expansion, resolution of the magnetic field data etc can play a significant role in this process. Consequently, the footpoints of the loops may be rooted (in the photospheric layer) in locations different from where they appear in the coronal images. Nonetheless, in all cases of Fig.~\ref{fig:context_loop1}, we observe oscillatory motions accompanying such `split-drift' events. In the following section we discuss about those oscillations.

\subsection{Presence of kink oscillations}
\begin{figure}[!ht]
\centering
\includegraphics[width=0.49\textwidth,clip,trim=0cm 1cm 0cm 0cm]{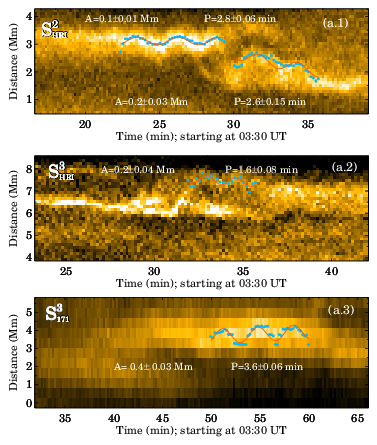}
 \caption{Zoomed-in views of the kink oscillations occurring in the loops. Panels a.1, a.2, and a.3 display portions of the x-t maps derived from slits \slithri{2}, \slithri{3}, and \slitaia[3]{171}, respectively. In each x-t map, the cyan dots, as determined by fitting a Gaussian function across the transverse direction of these maps, indicate the outline of the oscillating thread, while the red curve represents the fitted sinusoidal function to the detected thread. The amplitude and period values for each fitted thread are displayed on the panels.
}
\label{fig:oscillation}
\end{figure}

Figure \ref{fig:oscillation} provides zoomed-in views of three x-t maps taken from Fig~\ref{fig:context_loop1}. We observe signatures of transverse oscillations occurring not only after the loop's split but also before the split and, in some cases, during both situations. This raises questions about whether these oscillations are a cause of the observed splits or simply a response to the sudden separation events. Nonetheless, most of these oscillations appear to survive multiple cycles without noticeable decrease in their amplitudes\footnote{Even though the oscillation in Fig. 3(a.1) starting at 30 min is not as clear as the earlier oscillation in the same panel, the sinusoidal fit is still much better than, e.g., a linear fit.}. The derived amplitudes and period values, as shown in Fig~\ref{fig:oscillation}, are typical of decayless kink oscillations reported in the literature \citep{2015A&A...583A.136A,2022A&A...666L...2M}. This further adds complexity to our findings. For example, in loop-1, the low amplitude observed does not align with the expected wave amplitude if the wave is a result of footpoint movement, as discussed in the previous section (one would anticipate a significantly larger amplitude in that case given the exponential drop in density).   

\subsection{Cross-field drift speeds}

\begin{figure}[!ht]
\centering
\includegraphics[width=0.49\textwidth,clip,trim=0cm 2cm 0cm 0cm]{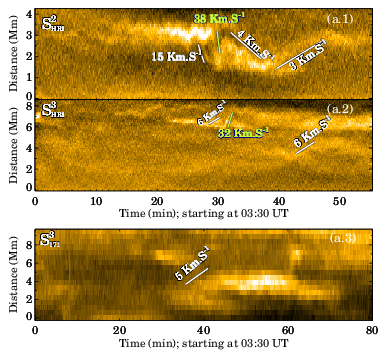}
 \caption{Estimation of drift speeds of the loops. The x-t maps displayed in the three panels are the same as those in Fig.~\ref{fig:oscillation}. To guide the eye, the white slanted lines outline the slopes that indicate the drift speeds of the loops, while the two green lines represent the slopes associated with the loop splits. The derived speed values are printed on the panels.}
\label{fig:speeds}
\end{figure}

As mentioned earlier, we observe two distinct types of cross-field movements during these events: i) movement of a thread-like structure that separates from the main loop; and ii) movement of the separated thread  towards its new location and later to its initial position. The animations associated with Fig~\ref{fig:context_loop1} demonstrate that these two speeds are notably different. We quantitatively establish this by measuring the slopes of the bright ridges of the x-t maps as presented in Fig.~\ref{fig:speeds}.  The sudden thread separation happens with speeds $\approx$30~km.s$^{-1}$ (slopes drawn with cyan lines), while those loops drift much slowly with speeds $\approx$5~km.s$^{-1}$ (slopes drawn with white lines). These speeds do not correspond to any characteristic velocities typically observed in the lower corona. However, since these speeds consistently appear across all events, it is reasonable to think they are associated with the driver that triggers these `split-drift' cases. One possibility is that these speeds reflect the motion of the loop footpoints, which go along with the granular flow that is typically found to be 1-2~km.s$^{-1}$. However, in most cases, the drift speeds are 2-3 times higher than these values. As presented in Section \ref{sec:magnetic_field}, the magnetic patches at the photosphere do not seem to move significantly and therefore, these speeds likely do not have a direct connection to the photosphere. Still, considering the strong density gradient between the photosphere and the corona, it is possible that the relatively slow drift speeds in the photosphere could lead to much faster changes in the upper atmosphere. However, this effect may be limited to scenarios involving faster footpoint driving (leading to waves) and may not apply to slower driving associated with quasi-static evolution.

Additionally, chromospheric and transition region features, such as dynamic fibrils or spicules, exhibit characteristic speeds of 15-30~km~s$^{-1}$ \citep{2023A&A...670L...3M}, which resemble the drift speeds observed in this work. Even so, it remains unclear how the field-aligned speeds of spicules or dynamic fibrils are related to the transverse (to the field) drift speeds of the observed loops. Additionally, spicules also exhibit transversal motions with wave amplitudes of about 20\,km\,s$^{-1}$ \citep{2007PASJ...59S.655D}. How such chromospheric oscillations could lead to split-drift motions that we observed closer to the coronal apexes of loops (e.g., slits 1, 2, and 4 in Fig.\,\ref{fig:context_loop1}) is unclear. 
Additionally, small-scale interchange reconnection at the footpoints of these loops can be a potential source. The loops we studied are anchored in weak plage or network structures that are located away from the core of the active region. Here, magnetoconvection coupled with dynamic parasitic polarity activity on short timescales of less than 5\,minutes could create disturbances in the corona \citep[][]{2023ApJ...956L...1C}. While HMI magnetograms do indicate some evidence of scattered parasitic polarities near the footpoints of the coronal loops (Fig.\,\ref{fig:hmi_171}), we cannot establish a clear one-to-one correspondance between potential small-scale interchange reconnection and the observed coronal loop drifts. 

\section{Summary and Discussion}
Let us first summarize the key features of these `split-drift' events:

1. Two characteristic speeds are involved: a faster (30\,km\,s$^{-1}$) away movement of a thin, thread-like structure and a substantially slower (5\,km\,s$^{-1}$) drift speed of the parent or detached loop.
   
2. Kink oscillations are present either before or after the split, and sometimes both.

3. The back-and-forth movement of one of the loop footpoints is evident in EUV coronal images. Line-of-sight (LOS) magnetogram data confirms the presence of two similar polarity patches where the footpoints shift. However, the magnetic patches (in the HMI data) themselves do not move in the process.

4. The evolution of these events in cooler channels (such as in AIA~\aia{171} or \hrieuv) do not always correspond to what is observed in hotter channels (such as in AIA~\aia{193} or the \aia{211} passbands).

Several aspects of the observed loop evolution remain challenging to interpret. For instance, in loop-1, the evolution captured in different AIA channels appears completely different with respect to the loop's position. This observation is inconsistent with the typical multithermal loop model \citep{2009ApJ...691..503S}, which would suggest some similarities in the evolution patterns between the AIA 171 and AIA 193 maps, as these two passbands share overlaps in their response functions. One possible explanation is that there are multiple (at least three) overlapping strands with specific temperature profiles that cause them to selectively appear or disappear in a particular AIA passband. The AIA \aia{193} image in Fig. \ref{fig:context_loop1}c indeed shows two overlapping strands at the slit-1 location, potentially explaining the apparent stationary appearance of one strand in that slit. However, requiring such specific temperature profiles seems somewhat contrived and thus less likely. On the other hand, a closer look at the AIA data hints at the multithermal nature of these strands. The intensity of a prominent bright loop, which remains stationary in the AIA \aia{193} map (Fig. \ref{fig:context_loop1}b.2), significantly decreases right after the split is observed in the AIA \aia{171} map, further suggesting multithermality. The long-term loop intensity evolution discussed in \lastpaper~adds additional evidence for this. Still, the rapid change in a strand’s appearance across two AIA passbands within a few minutes remains puzzling.
 In addition, in Fig.~\ref{fig:context_loop1}a.2, we find enhanced emission prior to the split, which could be interpreted as a signature of localized heating (or cooling). However, as outlined in Section 3.2, the evolution in the hotter channels appears quite different and does not conform to a single scenario. Furthermore, in other events, such intensity enhancements are not immediately apparent. Therefore, at this point, we do not have conclusive evidence of localized heating or cooling events.

Another intriguing aspect is the lack of kink oscillation signals in the AIA \aia{193} and AIA \aia{211} maps, despite their presence in AIA \aia{171} map. As mentioned earlier, the oscillation parameters observed are characteristic of decayless kink oscillations reported in the literature. However, unlike typical kink oscillations, which usually appear across multiple AIA passbands \citep{2012ApJ...751L..27W,2021A&A...652L...3M}, the oscillations in this study do not. This discrepancy may support the idea of a multi-strand loop structure with unusual temperature profiles. The absence of oscillation signals in the hotter passbands could also be attributed to the diffuse nature of the hotter plasma \citep{2009ApJ...694.1256T}, which may obscure oscillation signatures. However, this trend is not universal across maps; for example, the loops in \slitaia[3]{193} and \slitaia[4]{193} appear relatively compact, and therefore, once again, a coherent picture does not emerge from all the events analyzed.

The pattern of speeds with which the loop drifts seems to be consistent across events (Fig.~\ref{fig:speeds}). As discussed in Section~\ref{sec:magnetic_field}, the magnetic patches in the photosphere near one of the loop footpoints remain stationary, while the EUV footpoint exhibits significant back-and-forth movement. This observation suggests that the origin of this movement likely occurs in the layers above the photosphere. As shown in Fig~\ref{fig:appendix_multiwave}e, the cooler \aia{304} passband data do not indicate the presence of low-lying filament-like structures. Therefore, the height of the perturbation, as well as its nature, which drives these 'split-drift' events, remains unclear at this point. Additionally, as mentioned in Section~\ref{sec:magnetic_field}, we should exercise caution when associating the footpoints derived from EUV images with their counterparts in the lower atmosphere.

Considering the points we have discussed thus far, it appears that magnetic reconnection could be a plausible source driving these `split-drift' events. The rapid timescale of these splits (occurring in under a minute) and the oscillations observed after the split agree well with the idea that these events result from magnetic reconnection. Conceptually, it is straightforward to imagine two slightly misaligned magnetic field lines reconnecting through the `component reconnection' mechanism, forming structures that drift away perpendicularly \citep{2021A&A...656A.141P}. On top of that, if we invoke such reconnections at different locations along a strand (caused by multiple other strands), the resulting effect can easily explain the return of the loops. According to this picture, the observed loop motions relate to the actual motion of the field lines. However, this scenario is not without caveats. For example, none of the cases analyzed here exhibited the typical outflowing jet-like features (`nanojets') commonly associated with such reconnection events \citep{2021NatAs...5...54A}, nor do they show sudden brightenings associated with untangling of braided loops \citep[][]{2022A&A...667A.166C}. Furthermore, the heating associated with reconnection is generally impulsive in nature, and it takes considerable time (ca 1000 seconds) for the emission to become visible in most observing channels \citep{2008ApJ...682.1351K}.

Nevertheless, this does not rule out the reconnection hypothesis. Reconnected field lines demonstrate real motion near the reconnection site, where the outflows can be extremely fast. However, as we move away from the site, this motion becomes limited and disappears entirely at the line-tied footpoints, except for unrelated, slower photospheric flows at approximately 1\,km\,s$^{-1}$. In contrast to these real motions, there are apparent motions caused by the successive heating and cooling of adjacent field lines. Reconnected field lines become bright when the hot evaporated plasma that results from impulsive heating cools through the range of temperature sensitivity of the observing channel. The rate at which this effect propagates perpendicular to the field – which is the speed of apparent motion – is determined by that rate at which magnetic flux is processed by the reconnection. Typically, reconnection inflow velocities are around 1\% of the local Alfv\'{e}n speed\footnote{Although we do not have direct measurements of the magnetic field and the density of the loop, it is reasonable to assume typical coronal values for these parameters. This assumption would suggest a local Alfv\'{e}n speed between 1000 and 3000\,km\,s$^{-1}$.} based on the reconnecting (shear) field component, or roughly 10 to 30\,km\,s$^{-1}$. Thus, the apparent drift of a bright strand should be comparable to this speed well away from the reconnection site, which is consistent with what we observe. We note that an apparent drift associated with cooling would initially appear in the \aia{211} and \aia{193} channels, followed by the \aia{171} channel, which is consistent with the observations. In this picture, however, it is challenging to understand how the thread returns to its initial position as seen in some of our events.  

Another type of apparent motion related to successive heating and cooling is associated with phenomena known as "magnetic flipping" \citep{2003JGRA..108.1285P,2005PhPl...12e2307P} and "slip-running" reconnection \citet{2006SoPh..238..347A}. The speeds in these cases can be very fast—potentially even super-Alfvenic—depending on the thickness of the reconnecting current sheet. These apparent motions occur parallel to the current sheet, while the motions discussed earlier are perpendicular. There are a couple of important questions that arise from the discussion above: (i) Given that such observations are rare in coronal loops (the events discussed in \citealp{2007Sci...318.1588A} share some similarities with the events addressed here), what conditions need to be met for these occurrences to be observed more frequently in coronal loops? (ii) What factors, including projection effects, influence whether the apparent speeds are sub- or super-Alfv\'{e}nic? Unfortunately, the data we have for these events is insufficient to answer these questions. Therefore, in the future, we need to conduct a statistical study of loops at different positions on the solar disc. Additionally, examining a realistic multi-strand loop model, such as the one presented in \cite{2022A&A...658A..45B}, may help clarify the role of reconnection in producing the observed dynamics in these loops.

\section{Conclusion}
In this paper, we presented several examples of unusual, cross-field motions of coronal loops. These events are characterized by the presence of kink oscillations both before and after the occurrences, along with distinct fast (30~km.s$^{-1}$) and slow (5~km.s$^{-1}$) cross-field drift speeds. Additionally, there is a noticeable back-and-forth movement of the loop footpoints between magnetic polarity patches. While we discussed several possible causes including magnetic reconnection that are likely to be driving these changes, a comprehensive understanding of these events is still lacking. To gain deeper insight into this phenomenon, future studies with larger statistical samples are necessary.
\begin{acknowledgements}
We thank the anonymous reviewer for their insightful comments. The EUI instrument was built by CSL, IAS, MPS, MSSL/UCL, PMOD/WRC, ROB, LCF/IO with funding from the Belgian Federal Science Policy Office (BELSPO/PRODEX PEA 4000112292 and 4000134088); the Centre National d’Etudes Spatiales (CNES); the UK Space Agency (UKSA); the Bundesministerium für Wirtschaft und Energie (BMWi) through the Deutsches Zentrum für Luft- und Raumfahrt (DLR); and the Swiss Space Office (SSO). We are grateful to the ESA SOC and MOC teams for their support. Solar Dynamics Observatory (SDO) is the first mission to be launched for NASA's Living With a Star (LWS) Program. The data from the SDO/AIA consortium are provided by the Joint Science Operations Center (JSOC) Science Data Processing at Stanford University. The work of S.M. is funded by the Federal Ministry for Economic Affairs and Climate Action (BMWK) through the German Space Agency at DLR based on a decision of the German Bundestag (Funding code: 50OU2201).
 L.P.C. gratefully acknowledges funding by the European Union (ERC, ORIGIN, 101039844). Views and opinions expressed are however those of the author(s) only and do not necessarily reflect those of the European Union or the European Research Council. Neither the European Union nor the granting authority can be held responsible for them. The work of JAK was supported by the GSFC Heliophysics Internal Scientist Funding Model competitive work package program. 
\end{acknowledgements}

\bibliography{loop_split_ref}

\begin{thebibliography}{40}
\expandafter\ifx\csname natexlab\endcsname\relax\def\natexlab#1{#1}\fi

\bibitem[{{Anfinogentov} {et~al.}(2015){Anfinogentov}, {Nakariakov}, \&
  {Nistic{\`o}}}]{2015A&A...583A.136A}
{Anfinogentov}, S.~A., {Nakariakov}, V.~M., \& {Nistic{\`o}}, G. 2015, \aap,
  583, A136

\bibitem[{{Antolin} {et~al.}(2021){Antolin}, {Pagano}, {Testa}, {Petralia}, \&
  {Reale}}]{2021NatAs...5...54A}
{Antolin}, P., {Pagano}, P., {Testa}, P., {Petralia}, A., \& {Reale}, F. 2021,
  Nature Astronomy, 5, 54

\bibitem[{{Aschwanden} {et~al.}(2008{\natexlab{a}}){Aschwanden}, {Nitta},
  {Wuelser}, \& {Lemen}}]{2008ApJ...680.1477A}
{Aschwanden}, M.~J., {Nitta}, N.~V., {Wuelser}, J.-P., \& {Lemen}, J.~R.
  2008{\natexlab{a}}, \apj, 680, 1477

\bibitem[{{Aschwanden} {et~al.}(2008{\natexlab{b}}){Aschwanden}, {W{\"u}lser},
  {Nitta}, \& {Lemen}}]{2008ApJ...679..827A}
{Aschwanden}, M.~J., {W{\"u}lser}, J.-P., {Nitta}, N.~V., \& {Lemen}, J.~R.
  2008{\natexlab{b}}, \apj, 679, 827

\bibitem[{{Aulanier} {et~al.}(2007){Aulanier}, {Golub}, {DeLuca}, {Cirtain},
  {Kano}, {Lundquist}, {Narukage}, {Sakao}, \& {Weber}}]{2007Sci...318.1588A}
{Aulanier}, G., {Golub}, L., {DeLuca}, E.~E., {et~al.} 2007, Science, 318, 1588

\bibitem[{{Aulanier} {et~al.}(2006){Aulanier}, {Pariat}, {D{\'e}moulin}, \&
  {Devore}}]{2006SoPh..238..347A}
{Aulanier}, G., {Pariat}, E., {D{\'e}moulin}, P., \& {Devore}, C.~R. 2006,
  \solphys, 238, 347

\bibitem[{{Breu} {et~al.}(2022){Breu}, {Peter}, {Cameron}, {Solanki},
  {Przybylski}, {Rempel}, \& {Chitta}}]{2022A&A...658A..45B}
{Breu}, C., {Peter}, H., {Cameron}, R., {et~al.} 2022, \aap, 658, A45

\bibitem[{{Chitta} {et~al.}(2022){Chitta}, {Peter}, {Parenti}, {Berghmans},
  {Auch{\`e}re}, {Solanki}, {Aznar Cuadrado}, {Sch{\"u}hle}, {Teriaca},
  {Mandal}, {Barczynski}, {Buchlin}, {Harra}, {Kraaikamp}, {Long}, {Rodriguez},
  {Schwanitz}, {Smith}, {Verbeeck}, {Zhukov}, {Liu}, \&
  {Cheung}}]{2022A&A...667A.166C}
{Chitta}, L.~P., {Peter}, H., {Parenti}, S., {et~al.} 2022, \aap, 667, A166

\bibitem[{{Chitta} {et~al.}(2017){Chitta}, {Peter}, {Solanki}, {Barthol},
  {Gandorfer}, {Gizon}, {Hirzberger}, {Riethm{\"u}ller}, {van Noort}, {Blanco
  Rodr{\'\i}guez}, {Del Toro Iniesta}, {Orozco Su{\'a}rez}, {Schmidt},
  {Mart{\'\i}nez Pillet}, \& {Kn{\"o}lker}}]{2017ApJS..229....4C}
{Chitta}, L.~P., {Peter}, H., {Solanki}, S.~K., {et~al.} 2017, \apjs, 229, 4

\bibitem[{{Chitta} {et~al.}(2023){Chitta}, {Solanki}, {del Toro Iniesta},
  {Woch}, {Calchetti}, {Gandorfer}, {Hirzberger}, {Kahil}, {Valori}, {Orozco
  Su{\'a}rez}, {Strecker}, {Appourchaux}, {Volkmer}, {Peter}, {Mandal}, {Aznar
  Cuadrado}, {Teriaca}, {Sch{\"u}hle}, {Berghmans}, {Verbeeck}, {Zhukov}, \&
  {Priest}}]{2023ApJ...956L...1C}
{Chitta}, L.~P., {Solanki}, S.~K., {del Toro Iniesta}, J.~C., {et~al.} 2023,
  \apjl, 956, L1

\bibitem[{{de Pontieu} {et~al.}(2007){de Pontieu}, {McIntosh}, {Hansteen},
  {Carlsson}, {Schrijver}, {Tarbell}, {Title}, {Shine}, {Suematsu}, {Tsuneta},
  {Katsukawa}, {Ichimoto}, {Shimizu}, \& {Nagata}}]{2007PASJ...59S.655D}
{de Pontieu}, B., {McIntosh}, S., {Hansteen}, V.~H., {et~al.} 2007, \pasj, 59,
  S655

\bibitem[{{Feng} {et~al.}(2007){Feng}, {Inhester}, {Solanki}, {Wiegelmann},
  {Podlipnik}, {Howard}, \& {Wuelser}}]{2007ApJ...671L.205F}
{Feng}, L., {Inhester}, B., {Solanki}, S.~K., {et~al.} 2007, \apjl, 671, L205

\bibitem[{{Judge} {et~al.}(2024){Judge}, {Kleint}, \&
  {Kuckein}}]{2024ApJ...970..147J}
{Judge}, P.~G., {Kleint}, L., \& {Kuckein}, C. 2024, \apj, 970, 147

\bibitem[{{Kaiser} {et~al.}(2008){Kaiser}, {Kucera}, {Davila}, {St. Cyr},
  {Guhathakurta}, \& {Christian}}]{2008SSRv..136....5K}
{Kaiser}, M.~L., {Kucera}, T.~A., {Davila}, J.~M., {et~al.} 2008, \ssr, 136, 5

\bibitem[{{Klimchuk}(2000)}]{2000SoPh..193...53K}
{Klimchuk}, J.~A. 2000, \solphys, 193, 53

\bibitem[{{Klimchuk}(2006)}]{2006SoPh..234...41K}
{Klimchuk}, J.~A. 2006, \solphys, 234, 41

\bibitem[{{Klimchuk} \& {DeForest}(2020)}]{2020ApJ...900..167K}
{Klimchuk}, J.~A. \& {DeForest}, C.~E. 2020, \apj, 900, 167

\bibitem[{{Klimchuk} {et~al.}(1992){Klimchuk}, {Lemen}, {Feldman}, {Tsuneta},
  \& {Uchida}}]{1992PASJ...44L.181K}
{Klimchuk}, J.~A., {Lemen}, J.~R., {Feldman}, U., {Tsuneta}, S., \& {Uchida},
  Y. 1992, \pasj, 44, L181

\bibitem[{{Klimchuk} {et~al.}(2008){Klimchuk}, {Patsourakos}, \&
  {Cargill}}]{2008ApJ...682.1351K}
{Klimchuk}, J.~A., {Patsourakos}, S., \& {Cargill}, P.~J. 2008, \apj, 682, 1351

\bibitem[{{Kraaikamp} {et~al.}(2023){Kraaikamp}, {Gissot}, {Stegen}, {Mampaey},
  {Verbeeck}, {Auch{\`e}re}, \& {Berghmans}}]{euidatarelease6}
{Kraaikamp}, E., {Gissot}, S., {Stegen}, K., {et~al.} 2023, SolO/EUI Data
  Release 6.0 2023-01, https://doi.org/10.24414/z818-4163, published by Royal
  Observatory of Belgium (ROB)

\bibitem[{{Lemen} {et~al.}(2012){Lemen}, {Title}, {Akin}, {Boerner}, {Chou},
  {Drake}, {Duncan}, {Edwards}, {Friedlaender}, {Heyman}, {Hurlburt}, {Katz},
  {Kushner}, {Levay}, {Lindgren}, {Mathur}, {McFeaters}, {Mitchell}, {Rehse},
  {Schrijver}, {Springer}, {Stern}, {Tarbell}, {Wuelser}, {Wolfson}, {Yanari},
  {Bookbinder}, {Cheimets}, {Caldwell}, {Deluca}, {Gates}, {Golub}, {Park},
  {Podgorski}, {Bush}, {Scherrer}, {Gummin}, {Smith}, {Auker}, {Jerram},
  {Pool}, {Soufli}, {Windt}, {Beardsley}, {Clapp}, {Lang}, \&
  {Waltham}}]{2012SoPh..275...17L}
{Lemen}, J.~R., {Title}, A.~M., {Akin}, D.~J., {et~al.} 2012, \solphys, 275, 17

\bibitem[{{Mandal} {et~al.}(2022){Mandal}, {Chitta}, {Antolin}, {Peter},
  {Solanki}, {Auch{\`e}re}, {Berghmans}, {Zhukov}, {Teriaca}, {Cuadrado},
  {Sch{\"u}hle}, {Parenti}, {Buchlin}, {Harra}, {Verbeeck}, {Kraaikamp},
  {Long}, {Rodriguez}, {Pelouze}, {Schwanitz}, {Barczynski}, \&
  {Smith}}]{2022A&A...666L...2M}
{Mandal}, S., {Chitta}, L.~P., {Antolin}, P., {et~al.} 2022, \aap, 666, L2

\bibitem[{{Mandal} {et~al.}(2023){Mandal}, {Peter}, {Chitta}, {Cuadrado},
  {Sch{\"u}hle}, {Teriaca}, {Solanki}, {Harra}, {Berghmans}, {Auch{\`e}re},
  {Parenti}, {Zhukov}, {Buchlin}, {Verbeeck}, {Kraaikamp}, {Rodriguez}, {Long},
  {Schwanitz}, {Barczynski}, {Pelouze}, {Smith}, {Liu}, \&
  {Cheung}}]{2023A&A...670L...3M}
{Mandal}, S., {Peter}, H., {Chitta}, L.~P., {et~al.} 2023, \aap, 670, L3

\bibitem[{{Mandal} {et~al.}(2024){Mandal}, {Peter}, {Klimchuk}, {Solanki},
  {Chitta}, {Aznar Cuadrado}, {Sch{\"u}hle}, {Teriaca}, {Berghmans},
  {Verbeeck}, {Auch{\`e}re}, \& {Stegen}}]{2024A&A...682L...9M}
{Mandal}, S., {Peter}, H., {Klimchuk}, J.~A., {et~al.} 2024, \aap, 682, L9

\bibitem[{{Mandal} {et~al.}(2021){Mandal}, {Tian}, \&
  {Peter}}]{2021A&A...652L...3M}
{Mandal}, S., {Tian}, H., \& {Peter}, H. 2021, \aap, 652, L3

\bibitem[{{McCarthy} {et~al.}(2021){McCarthy}, {Longcope}, \&
  {Malanushenko}}]{2021ApJ...913...56M}
{McCarthy}, M.~I., {Longcope}, D.~W., \& {Malanushenko}, A. 2021, \apj, 913, 56

\bibitem[{{M{\"u}ller} {et~al.}(2020){M{\"u}ller}, {St. Cyr}, {Zouganelis},
  {Gilbert}, {Marsden}, {Nieves-Chinchilla}, {Antonucci}, {Auch{\`e}re},
  {Berghmans}, {Horbury}, {Howard}, {Krucker}, {Maksimovic}, {Owen}, {Rochus},
  {Rodriguez-Pacheco}, {Romoli}, {Solanki}, {Bruno}, {Carlsson}, {Fludra},
  {Harra}, {Hassler}, {Livi}, {Louarn}, {Peter}, {Sch{\"u}hle}, {Teriaca}, {del
  Toro Iniesta}, {Wimmer-Schweingruber}, {Marsch}, {Velli}, {De Groof},
  {Walsh}, \& {Williams}}]{2020A&A...642A...1M}
{M{\"u}ller}, D., {St. Cyr}, O.~C., {Zouganelis}, I., {et~al.} 2020, \aap, 642,
  A1

\bibitem[{{Pagano} {et~al.}(2021){Pagano}, {Antolin}, \&
  {Petralia}}]{2021A&A...656A.141P}
{Pagano}, P., {Antolin}, P., \& {Petralia}, A. 2021, \aap, 656, A141

\bibitem[{{Pesnell} {et~al.}(2012){Pesnell}, {Thompson}, \&
  {Chamberlin}}]{2012SoPh..275....3P}
{Pesnell}, W.~D., {Thompson}, B.~J., \& {Chamberlin}, P.~C. 2012, \solphys,
  275, 3

\bibitem[{{Peter} \& {Bingert}(2012)}]{2012A&A...548A...1P}
{Peter}, H. \& {Bingert}, S. 2012, \aap, 548, A1

\bibitem[{{Peter} {et~al.}(2013){Peter}, {Bingert}, {Klimchuk}, {de Forest},
  {Cirtain}, {Golub}, {Winebarger}, {Kobayashi}, \&
  {Korreck}}]{2013A&A...556A.104P}
{Peter}, H., {Bingert}, S., {Klimchuk}, J.~A., {et~al.} 2013, \aap, 556, A104

\bibitem[{{Pontin} {et~al.}(2005){Pontin}, {Galsgaard}, {Hornig}, \&
  {Priest}}]{2005PhPl...12e2307P}
{Pontin}, D.~I., {Galsgaard}, K., {Hornig}, G., \& {Priest}, E.~R. 2005,
  Physics of Plasmas, 12, 052307

\bibitem[{{Priest} {et~al.}(2003){Priest}, {Hornig}, \&
  {Pontin}}]{2003JGRA..108.1285P}
{Priest}, E.~R., {Hornig}, G., \& {Pontin}, D.~I. 2003, Journal of Geophysical
  Research (Space Physics), 108, 1285

\bibitem[{{Reale}(2014)}]{2014LRSP...11....4R}
{Reale}, F. 2014, Living Reviews in Solar Physics, 11, 4

\bibitem[{{Rochus} {et~al.}(2020){Rochus}, {Auch{\`e}re}, {Berghmans}, {Harra},
  {Schmutz}, {Sch{\"u}hle}, {Addison}, {Appourchaux}, {Aznar Cuadrado},
  {Baker}, {Barbay}, {Bates}, {BenMoussa}, {Bergmann}, {Beurthe}, {Borgo},
  {Bonte}, {Bouzit}, {Bradley}, {B{\"u}chel}, {Buchlin}, {B{\"u}chner},
  {Cab{\'e}}, {Cadiergues}, {Chaigneau}, {Chares}, {Choque Cortez}, {Coker},
  {Condamin}, {Coumar}, {Curdt}, {Cutler}, {Davies}, {Davison}, {Defise}, {Del
  Zanna}, {Delmotte}, {Delouille}, {Dolla}, {Dumesnil}, {D{\"u}rig}, {Enge},
  {Fran{\c{c}}ois}, {Fourmond}, {Gillis}, {Giordanengo}, {Gissot}, {Green},
  {Guerreiro}, {Guilbaud}, {Gyo}, {Haberreiter}, {Hafiz}, {Hailey}, {Halain},
  {Hansotte}, {Hecquet}, {Heerlein}, {Hellin}, {Hemsley}, {Hermans}, {Hervier},
  {Hochedez}, {Houbrechts}, {Ihsan}, {Jacques}, {J{\'e}r{\^o}me}, {Jones},
  {Kahle}, {Kennedy}, {Klaproth}, {Kolleck}, {Koller}, {Kotsialos},
  {Kraaikamp}, {Langer}, {Lawrenson}, {Le Clech'}, {Lenaerts}, {Liebecq},
  {Linder}, {Long}, {Mampaey}, {Markiewicz-Innes}, {Marquet}, {Marsch},
  {Matthews}, {Mazy}, {Mazzoli}, {Meining}, {Meltchakov}, {Mercier}, {Meyer},
  {Monecke}, {Monfort}, {Morinaud}, {Moron}, {Mountney}, {M{\"u}ller},
  {Nicula}, {Parenti}, {Peter}, {Pfiffner}, {Philippon}, {Phillips},
  {Plesseria}, {Pylyser}, {Rabecki}, {Ravet-Krill}, {Rebellato}, {Renotte},
  {Rodriguez}, {Roose}, {Rosin}, {Rossi}, {Roth}, {Rouesnel}, {Roulliay},
  {Rousseau}, {Ruane}, {Scanlan}, {Schlatter}, {Seaton}, {Silliman}, {Smit},
  {Smith}, {Solanki}, {Spescha}, {Spencer}, {Stegen}, {Stockman}, {Szwec},
  {Tamiatto}, {Tandy}, {Teriaca}, {Theobald}, {Tychon}, {van Driel-Gesztelyi},
  {Verbeeck}, {Vial}, {Werner}, {West}, {Westwood}, {Wiegelmann}, {Willis},
  {Winter}, {Zerr}, {Zhang}, \& {Zhukov}}]{2020A&A...642A...8R}
{Rochus}, P., {Auch{\`e}re}, F., {Berghmans}, D., {et~al.} 2020, \aap, 642, A8

\bibitem[{{Scherrer} {et~al.}(2012){Scherrer}, {Schou}, {Bush}, {Kosovichev},
  {Bogart}, {Hoeksema}, {Liu}, {Duvall}, {Zhao}, {Title}, {Schrijver},
  {Tarbell}, \& {Tomczyk}}]{2012SoPh..275..207S}
{Scherrer}, P.~H., {Schou}, J., {Bush}, R.~I., {et~al.} 2012, \solphys, 275,
  207

\bibitem[{{Schmelz} {et~al.}(2009){Schmelz}, {Nasraoui}, {Rightmire}, {Kimble},
  {del Zanna}, {Cirtain}, {DeLuca}, \& {Mason}}]{2009ApJ...691..503S}
{Schmelz}, J.~T., {Nasraoui}, K., {Rightmire}, L.~A., {et~al.} 2009, \apj, 691,
  503

\bibitem[{{Solanki} {et~al.}(2020){Solanki}, {del Toro Iniesta}, {Woch},
  {Gandorfer}, {Hirzberger}, {Alvarez-Herrero}, {Appourchaux}, {Mart{\'\i}nez
  Pillet}, {P{\'e}rez-Grande}, {Sanchis Kilders}, {Schmidt}, {G{\'o}mez Cama},
  {Michalik}, {Deutsch}, {Fernandez-Rico}, {Grauf}, {Gizon}, {Heerlein},
  {Kolleck}, {Lagg}, {Meller}, {M{\"u}ller}, {Sch{\"u}hle}, {Staub}, {Albert},
  {Alvarez Copano}, {Beckmann}, {Bischoff}, {Busse}, {Enge}, {Frahm},
  {Germerott}, {Guerrero}, {L{\"o}ptien}, {Meierdierks}, {Oberdorfer},
  {Papagiannaki}, {Ramanath}, {Schou}, {Werner}, {Yang}, {Zerr}, {Bergmann},
  {Bochmann}, {Heinrichs}, {Meyer}, {Monecke}, {M{\"u}ller}, {Sperling},
  {{\'A}lvarez Garc{\'\i}a}, {Aparicio}, {Balaguer Jim{\'e}nez}, {Bellot
  Rubio}, {Cobos Carracosa}, {Girela}, {Hern{\'a}ndez Exp{\'o}sito}, {Herranz},
  {Labrousse}, {L{\'o}pez Jim{\'e}nez}, {Orozco Su{\'a}rez}, {Ramos},
  {Barandiar{\'a}n}, {Bastide}, {Campuzano}, {Cebollero}, {D{\'a}vila},
  {Fern{\'a}ndez-Medina}, {Garc{\'\i}a Parejo}, {Garranzo-Garc{\'\i}a},
  {Laguna}, {Mart{\'\i}n}, {Navarro}, {N{\'u}{\~n}ez Peral}, {Royo},
  {S{\'a}nchez}, {Silva-L{\'o}pez}, {Vera}, {Villanueva}, {Fourmond}, {de
  Galarreta}, {Bouzit}, {Hervier}, {Le Clec'h}, {Szwec}, {Chaigneau},
  {Buttice}, {Dominguez-Tagle}, {Philippon}, {Boumier}, {Le Cocguen},
  {Baranjuk}, {Bell}, {Berkefeld}, {Baumgartner}, {Heidecke}, {Maue}, {Nakai},
  {Scheiffelen}, {Sigwarth}, {Soltau}, {Volkmer}, {Blanco Rodr{\'\i}guez},
  {Domingo}, {Ferreres Sabater}, {Gasent Blesa}, {Rodr{\'\i}guez
  Mart{\'\i}nez}, {Osorno Caudel}, {Bosch}, {Casas}, {Carmona}, {Herms},
  {Roma}, {Alonso}, {G{\'o}mez-Sanjuan}, {Piqueras}, {Torralbo}, {Fiethe},
  {Guan}, {Lange}, {Michel}, {Bonet}, {Fahmy}, {M{\"u}ller}, \&
  {Zouganelis}}]{2020A&A...642A..11S}
{Solanki}, S.~K., {del Toro Iniesta}, J.~C., {Woch}, J., {et~al.} 2020, \aap,
  642, A11

\bibitem[{{Tripathi} {et~al.}(2009){Tripathi}, {Mason}, {Dwivedi}, {del Zanna},
  \& {Young}}]{2009ApJ...694.1256T}
{Tripathi}, D., {Mason}, H.~E., {Dwivedi}, B.~N., {del Zanna}, G., \& {Young},
  P.~R. 2009, \apj, 694, 1256

\bibitem[{{Wang} {et~al.}(2012){Wang}, {Ofman}, {Davila}, \&
  {Su}}]{2012ApJ...751L..27W}
{Wang}, T., {Ofman}, L., {Davila}, J.~M., \& {Su}, Y. 2012, \apjl, 751, L27

\end{thebibliography}
\bibliographystyle{aa}

\begin{appendix}

\section{Overall magnetic configuration}

\begin{figure*}[!ht]
\centering
\includegraphics[width=0.90\textwidth,clip,trim=0cm 0cm 0cm 0cm]{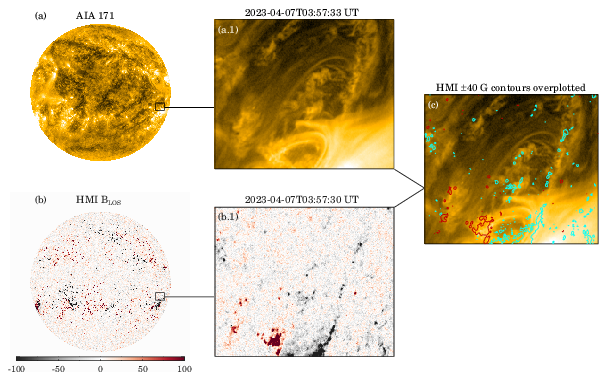}
 \caption{Images of the solar disc as recorded through the AIA~\aia{171} passband and the HMI-LOS ares shown in panel a and panel b, respectively. The black rectangle in each panel outlines the region of interest (ROI) that encompasses the loop system studied in this paper. Panels a.1 and b.1 provide additional images of the ROI for the AIA and HMI data, respectively. In panel c, the cyan and red contours, representing the boundaries of $\pm$40~G as derived from panel b.1, are overlaid on the AIA image from panel a.1.}
\label{fig:appendix_context}
\end{figure*}

In Figure~\ref{fig:appendix_context}, we provide an overview of the magnetic structure adjacent to the loop system studied in this work. As shown in panel b, the loop system is located away from an active region, where the magnetic field is relatively weaker near both of its footpoints. The footpoints on the west side are anchored within negative polarity patches, while the footpoints on the east side (not fully visible) are anchored in positive polarity patches.

\begin{figure}[!ht]
\centering
\includegraphics[width=0.49\textwidth,clip,trim=0cm 0cm 0cm 0cm]{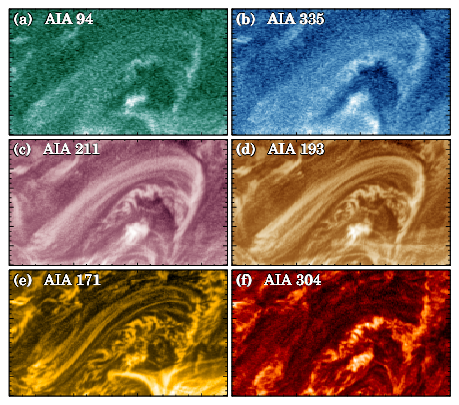}
 \caption{Sanpshots from different AIA passbands of the same FOV as shown in Fig.~\ref{fig:context_loop1}. }
\label{fig:appendix_multiwave}
\end{figure}

We also include Fig.~\ref{fig:appendix_multiwave} that presents images from different AIA passbands. This figure provides an overview of the multi-thermal structuring of the entire region. For example, in the \aia{94} snapshot, we observe very faint emissions from the overarching loops, which are likely much hotter. In contrast, these emissions are entirely absent from the cooler chromospheric (and transition region) emission captured in the \aia{304} emissions.

\end{appendix}

\end{document}